# Bright subcycle XUV pulse from a single dense relativistic electron sheet

W.J. Ma[1*], J.H. Bin[1,2], H.Y. Wang[2,3], M. Yeung[4,5], C. Kreuzer[1], M. Streeter[6], P.S. Foster[7], S. Cousens[4], D. Kiefer[1], B. Dromey[4], X.Q. Yan[3], M. Zepf[4,5], J. Meyer-ter-Vehn[2], J. Schreiber[1,2¶]

[1]Faculty of Physics, Ludwig-Maximilians-University, Munich, Germany,
[2]Max-Planck-Institute of Quantum Optics, Hans-Kopfermann-Str.1, D-85748 Garching, Germany,
[3]State Key Laboratory of Nuclear Physics and Technology, Peking University, Beijing 100871, China,
[4]Department of Physics and Astronomy, Queen's University Belfast, Belfast, BT7 1NN, UK,
[5]Helmholtz Institute Jena, 07743 Jena, Germany,
[6]Blackett Laboratory, Imperial College London, SW7 2BZ, UK,
[7]Central Laser Facility, STFC Rutherford Appleton Laboratory, Chilton, Didcot, OX11 0QX, UK

**Relativistic electrons are prodigious sources of photons. Beyond classical accelerators, ultra-intense laser interactions are of particular interest as they allow the coherent motion of relativistic electrons to be controlled and exploited as sources of radiation. Under extreme laser conditions theory predicts that isolated free relativistic electron sheets (FRES) can be produced[1] and exploited for the production of a new class of radiation – unipolar XUV pulses[2]. However, the combination of extremely rapid rise-time and highest peak intensity in these simulations is still beyond current laser technology. We demonstrate a route to isolated FRES with existing lasers by exploiting relativistic transparency[3] to produce an ultra-intense pulse with a steep risetime. When such an FRES interacts with a second, oblique target foil the electron sheet is rapidly accelerated ('kicked'). The radiative signature and simulations demonstrate that a single, nanometere thick FRES was produced. The experimental observations together with our theoretical modeling suggest the production of the first unipolar (half-cycle) pulse in the XUV – an achievement that has so far only been realized in the terahertz spectral domain.**

*email: Wenjun.Ma@physik.uni-muenchen.de
¶email: joerg.schreiber@mpq.mpg.de

Achieving such isolated, possibly even half-cycle, radiation bursts in the extreme ultraviolet (XUV) and X-Ray region are of great importance for time-resolved studies in atomic physics[4], solid state physics[5] and high-energy-density matter physics[6]. Currently the most brilliant XUV/X-Ray radiation facilities, for example free electron lasers (FEL), routinely offer bursts with duration down to hundreds of femtoseconds which can be further reduced to femtosecond scale by tailoring the bunches[7, 8]. The challenge in entering attosecond timescales is to reduce the electron bunch duration to below one femtosecond while reaching highest possible electron densities, which is limited to $10^{18}$ cm$^{-3}$ in modern accelerators[9]. Ultrashort and highly intense (I>$10^{18}$ W/cm$^2$) near infrared (NIR) laser pulses recently pave their way into this prosperous field, as they can drive very dense electron sheets at solid-density-plasmas. Their periodic relativistic motion results in a train of bright attosecond XUV/X-Ray bursts[10].

Such dense relativistic electron sheets (RES), with nanometer scale thickness and micrometer scale diameter, are excellent radiation sources, particularly when the electrons undergo coherent motion in an externally applied field such as an ultra intense laser. The resultant high degree of transverse and longitudinal/temporal coherence leads to bright emission[11]. Oscillating or multiple RES naturally result in a temporal pulsetrain. However, theoretical work has suggested that it is possible to produce a single free relativistic electron sheet (FRES) by irradiating a nanometer-thin foil with an ultraintense (I>$10^{21}$W/cm$^2$) few-cycle laser pulse[1] capable of overcoming the Coulomb force that binds the electrons to the ions in the foil ( the so-called 'blow out' regime). Such a single free electron sheet is superior to an oscillating RES at the plasma-vacuum boundary as it allows single burst of coherent radiation to be produced, for example by reflecting a short laser pulse from an FRES acting as a "flying mirror"[12, 13]. Another intriguing application, as theoretically illustrated very recently, is to exploit the strong transverse acceleration (or 'kick') a laser driven FRES experiences when the co-propagating laser pulse is separated off using an oblique reflector foil[2]. When the laser-dressed, relativistic electrons reach this second, slightly thicker, the laser field is reflected while the FRES propagates through the foil. Since the canonical momentum $p_x - a$ is is conserved, the electrons must respond to the sudden change of the dressing field configuration at the reflection point electron and the transverse momentum in laser polarization direction $p_x$ changes to zero in a fraction of an optical cylcle. This sudden change in field configuration results in a strong, single transverse kick and thus results in emission of a single burst of high-frequency radiation. Such asymmetric XUV-fields have many applications beyond previous realizations at terahertz frequencies[14] , such as the tracking and controlling of electron wave packets in atoms[15] or for optical-field-driven electronics[16] .

In spite of the appealing features of FRESs, generation of a single FRES in the 'blow-out' regime discussed in the literature to date is far beyond the capabilities of current laser technology, due to the difficulty of producing pulses that combine extremely sharp optical rise times with extreme intensities. In this paper, we overcome these limitations by exploiting relativistically induced transparency[3] to steepen a 50 fs pulse at peak intensity of 2x$10^{20}$ W/cm². The rapid rise of the steepened laser pulse results in increased ponderomotive force ($F_p \propto \nabla E^2$) and acts like a snowplough on the subsequent low density plasma and piles up electrons in one dominant electron density spike[17, 18] – thus forming a single FRES with comparable properties to those in the blow-out regime. Figure 1a,b represents this process schematically. Two-dimensional particle-in-cell simulations, performed for realistic laser and target parameters (Methods) are shown in figures 1c-f. Similarly to the 'blow-out' case discussed in the literature[12] the simulations show a free relativistic electron sheet sitting in the first well of the ponderomotive potential ('bucket') of the steepened pulse (figure 1c). The electrons that contribute to the FRES behave like free electrons in a laser field with normalized vector potential $a$ and consequently their transverse canonical momentum $p_x - a$ is conserved (fig. 1e). When the laser-dressed electrons approach the second foil, the electrons pass through whereas the laser is reflected. The expected switching of the electron transverse momentum in laser polarization direction from $p_{x0}$ to zero can be seen clearly (figure 1f). Since the reflected laser is now propagating along the y-axis the $v \times B$ force of results in the electrons gaining transverse momentum in y-direction.

In our simulation, the emerging electron sheet has a transverse diameter of 4 µm, a maximum density of $n_e$~$10^{21}$/cc (0.9$n_c$), and contains a charge of about 400 pC. The forward momenta peak at $p_z = 10mc$. On top of this relativistic motion in z-direction, the electrons have picked up a transverse momentum of $p_y = 1.1mc$, in good agreement with the theoretical prediction of $p_y \approx p_{x0}^2 tan\theta/2p_z \approx 1.3$ (see supplementary materials). Note that this momentum gain mainly happens when the first dominant 'bucket' is reflected and sweeps over the FRES, i.e., over one quarter of a laser wavelength ~200 nm. The electrons in the sheet are therefore deflected by $p_y/p_z$ ~ 100 mrad and must emit synchrotron radiation with a broad spectrum and critical frequency of ~148 eV, corresponding to a spike of $E_y$ in the time domain (XUV emission resulting from the synchrotron motion in $xz$ plane is trivial here due to the much higher corresponding critical frequency). Such a high-frequency field propagating with the FRES is observed in our simulation, but due to the high frequencies the corresponding attosecond scale temporal profile is not well resolved. Previous studies reveal the laser-to-XUV efficiency of such radiation can be as high as $10^{-4}$ and should therefore be easily detected thereby providing evidence for the presence and temporal structure of any FRES.

The radiation emitted from laser-irradiated double-foil targets was investigated experimentally. Free-standing diamond-like carbon (DLC) foils of 4 nm to 20 nm thick were used as single-foil targets or assembled as double-foil targets. They were irradiated by 50 fs laser pulses at peak intensity of $2\times10^{20}$W/cm$^2$. We were mainly interested in the emission of XUV-radiation, which was recorded angularly and spectrally resolved in the direction of main laser propagation (see figure 2a and Methods).

Figure 2a,b shows that over the observable spectral range from 10 to 400 eV the recorded spectra indeed strongly depend on target assembly and irradiation geometry. Note that the line emission remains similar for all cases and serves as calibration candle (see Methods). Regardless of the angle of incidence, single foil targets produce weak emission extending up to ~70 eV. The spectral modulations at multiples of the laser frequency between 10eV and 15 eV indicate the well-known emission of XUV-pulse trains due to the repetitive and periodic generation of electron bunches. When double foil targets are irradiated at normal incidence the signal is already stronger. But the emission changes dramatically when the double foil targets are tilted by an angle of 30° such that the laser is s-polarized with a bright XUV continuum extending up to 280 eV observable. Its spectral intensity is one order of magnitude larger over the complete spectral range accessible with the spectrometer and the absence of modulation indicates the presence of only one dominant radiation burst.

Meanwhile, the large intensity underlines the high degree of coherence of the emission process. As shown in Fig. 3, the spectral intensity of incoherent synchrotron radiation emitted by the FRES with a charge of 400 pC, as observed in our simulation, would be 6-7 order of magnitude lower than the experimentally registered spectra. The observed spectrum therefore resembles the coherently emitted part of the radiation. The corresponding spectral shape is then given by the form-factor, i.e. the longitudinal Fourier-transform, of the emitting electron sheet. One possible electron density distribution that reproduces the measured spectrum is compared to the simulation result in Fig. 3b and shows reasonable agreement (Methods). The absolute magnitude of the spectral intensity would suggest as much as 1000 pC of charge, again close to the value obtained from the simulations.

Further information as to the transverse coherence of the emission process can be obtained from the angular distribution of the observed radiation. Figure 2d shows the ratio of the off-axis intensity at 1.3° to on-axis intensity $I_{1.3°}/I_{0°}$ as a function of wavelength for the cases discussed before. On this angular range, only the bright, broadband emission from obliquely irradiated double-foil-targets shows clear evidence of directional emission. More specifically, the value $I_{1.3°}/I_{0°}$ increases from 0.65 to 0.78 with wavelength in the accessible range of 50 to 75 nm. This suggests that the emitted radiation has flat phase over an area with diameter of $d$=0.9 µm at the source (see Methods). This diameter determines the distance $L_d \sim d^2/c\tau$ over which any half-cycle pulse maintains its half-cycle nature before evolving into a single cycle pulse due to diffraction-induced transformation. For our parameters, $L_d \sim$670 µm is obtained which may limit the applications of the half-cycle XUV pulse (see Methods). But this limitation can be overcome to a certain extent by appropriate refocusing[19]. Future experimental challenges are characterizing the temporal structure of the XUV pulse and aiming to restore an asymmetric XUV field at some distance for temporal characterization and application of the pulse. Nevertheless, our studies represent the first step towards isolated, possibly half-cycle XUV pulses, which are very attractive for the studies in quantum information or petahertz electronics.

Already at this stage, the single burst with its large pulse energy of tens of µJ, small divergence of only 3.2° (56 mrad), and photon energies well beyond 100 eV highlight significant opportunities. The pulses are well suited to for single-shot attosecond-pump attosecond-probe experiments[20]. For physicists in other fields, such as high harmonic generation and coherent synchrotron emission, the case shown here provides a reference case, since the single unidirectional kick enables a complete understanding of the coherent emission process of an ultrathin electron bunch.

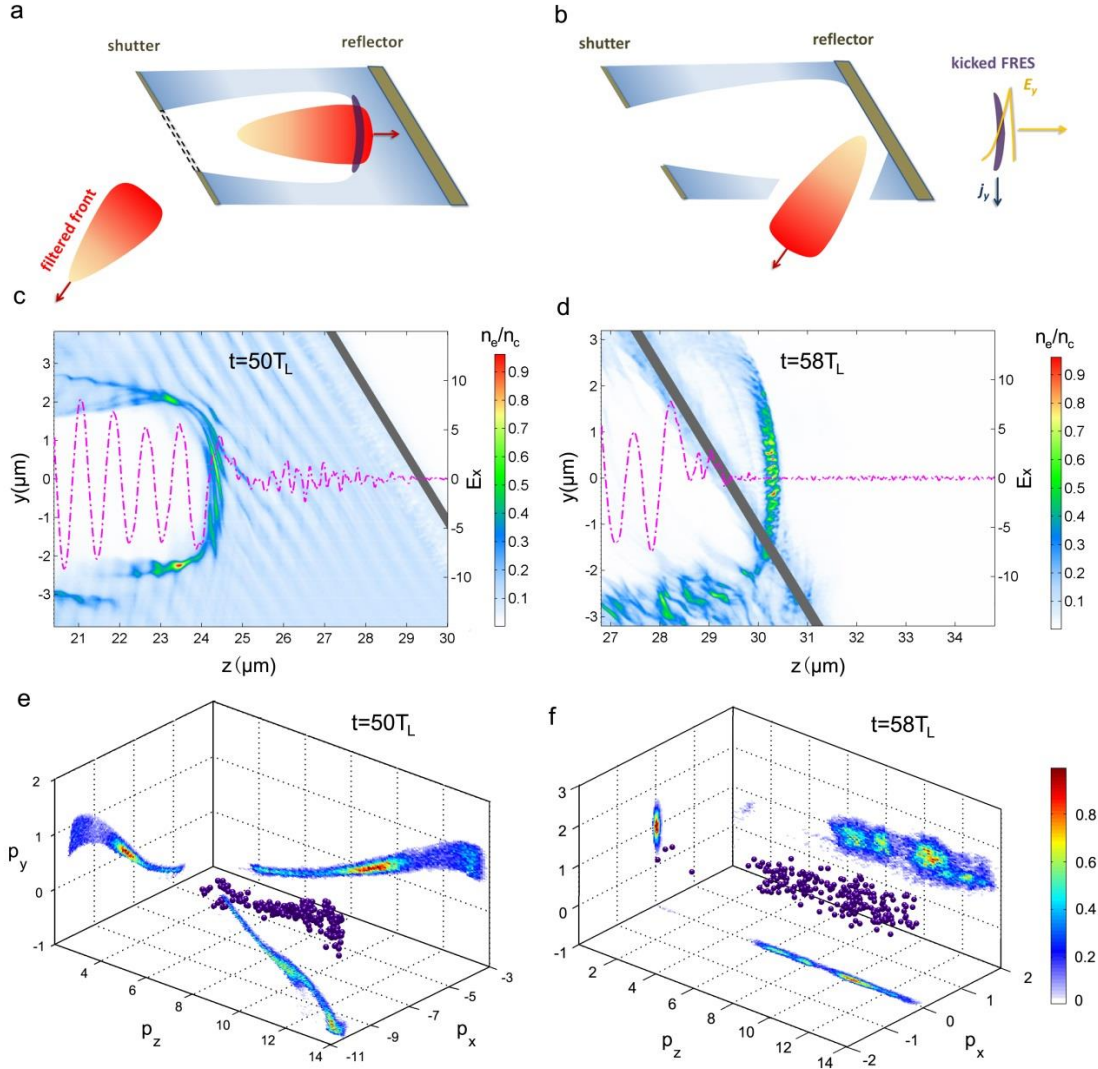

**Figure 1. Illustration of dense electron sheet generation in a tilted double foil target. a.** The front of a driving 50 fs laser pulse (red) is reflected by a nanometer thin foil (shutter), until the amplitude of the pulse is high enough to induce relativistic transparency. The pulse then breaks through the first foil with a very sharp rising edge, passes through dilute plasma (light blue), and piles up the free relativistic electron sheet (FRES, dark blue) in the first wave bucket. **b.** The second foil (reflector) is of key importance for shaking off the driving field from the FRES that passes the reflector keeping direction. Interaction with the sidewards reflected pulse switches electron momenta over a very short distance and leads to half-cycle XUV emission. **c,d.** Simulation results shown as snapshots in $y, z$ plane during sheet formation after 50 laser cycles ($t = 50 T_L$) and after shake-off ($t = 58 T_L$); electron density ($n_e/n_c$ according to color scale) and on-axis laser field ($eE_x/\hbar\omega c$, magenta) are presented. **e,f.** Phase space ($p_x, p_y, p_z$ in unit of $mc$) of FRES electrons corresponding to frames c,d, respectively. During shake-off of the laser field, electrons lose their quiver momenta ($p_x = 0$), but pick up uniform momentum in $y$ direction ($p_y = 1.1 mc$).

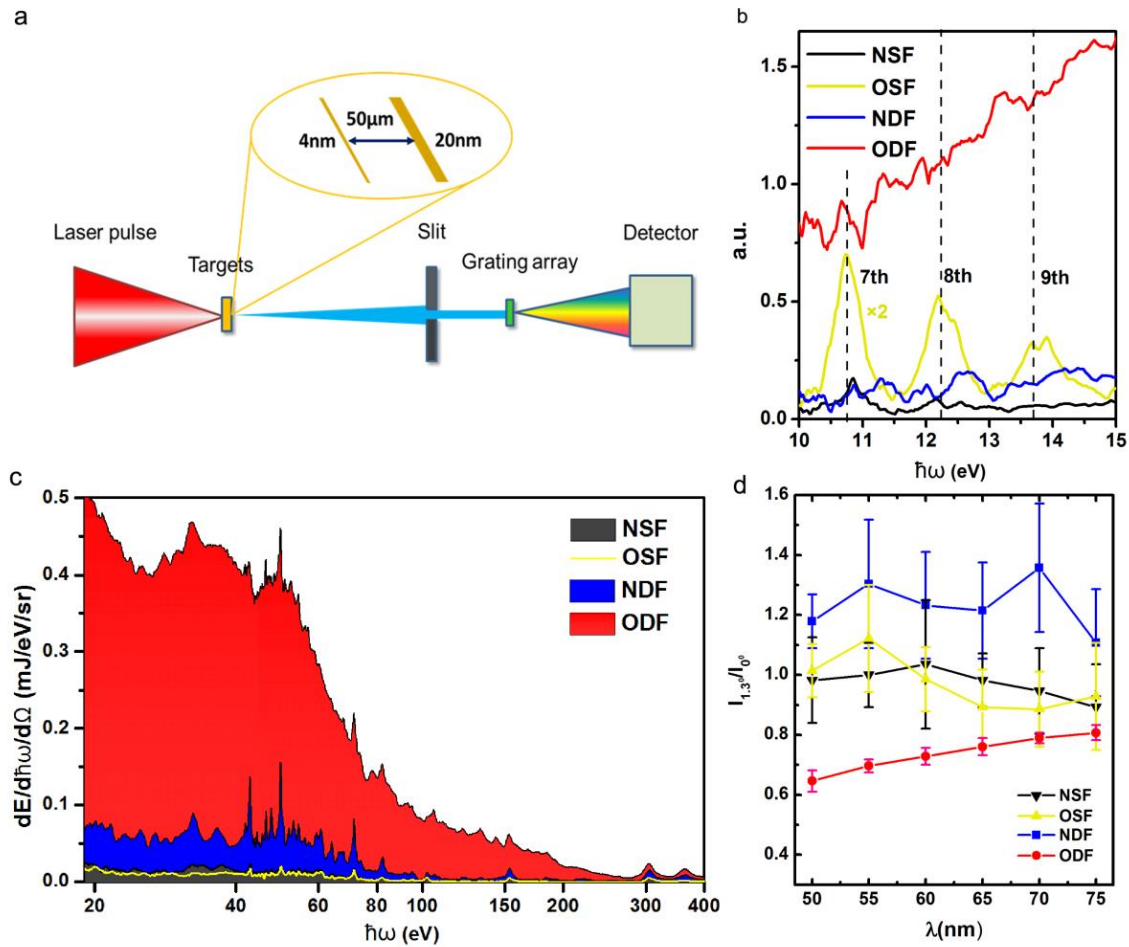

**Figure 2. . Experimental set-up and results**. **a.** S-polarized laser pulse (5J, 50 fs) is focused on 4 types of target configurations that differ in orientation and number of foils: normal single foil (NSF), oblique single foil (OSF), normal double foil (NDF), and oblique double foil (ODF). Single foils are made of 5 nm DLC, double foils are composed of 4 nm and 20 nm DLC placed 50 μm apart. XUV emission is spectrally resolved by a grating located behind a slit. **b**. The photon energy range of 5 – 15 eV is measured, using a 1,000 groove/mm transmission grating array. Dashed lines mark the $7^{th}, 8^{th}, 9^{th}$ harmonics; they are observed for the 5 nm single foil and indicate good contrast of the laser pulse. Since the spectra in this range are recorded behind a 60 nm Al filter without calibration, they give only relative amplitudes for different target geometry, but not the actual shape of the spectra. **c.** Spectra in the range of 18 - 400 eV measured with a 10,000 groove/mm transmission grating. **d.** Ratio of off-axis to on-axis intensity measured with the 1,000 groove/mm transmission grating array. The values are normalized to the case of OSF at 50 nm.

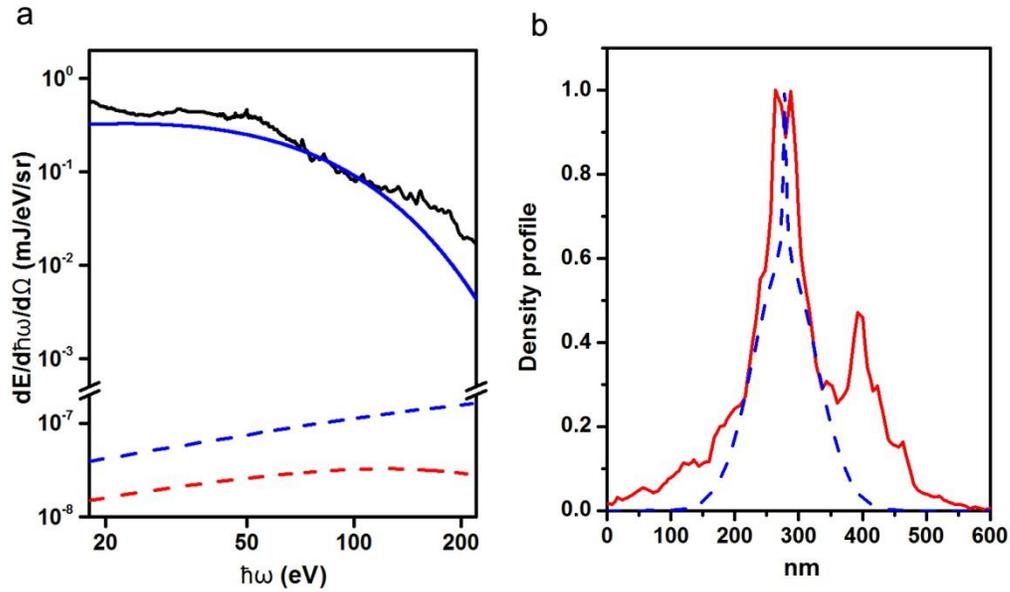

**Figure 3. Fitting of measured spectra with coherent synchrotron emission from a FRES.**
**a.** Black solid line gives the measured spectrum. For comparison, blue dashed line and red dashed line show incoherent synchrotron spectra calculated for 400 pC and 1000 pC electrons with $\gamma = 10$ and $\gamma = 15$, respectively. The blue solid line corresponds to coherent synchrotron emission, calculated for an electron sheet with a charge of 1000 pC, $\gamma = 15$, and the longitudinal profile shown in figure 3b. **b.** Blue dashed line is an assumed density profile which gives good fitting to the measured spectra in figure 3a. Red solid line shows the lineout of the simulated electron density along $y = 0$ at $t = 58T_L$.

**Methods**

**Experiment.** The experiment was performed at the ASTRA Gemini facility at Central Laser Facility of UK. This system delivers pulses with FWHM of 50 fs at central wavelength of 800 nm. A re-collimating double plasma mirror system was used to further enhance the contrast of the laser pulse to $10^{10}$ at -10 ps. An f/2 off-axis parabolic mirror was used to focus the pulse to an elliptical spot with FWHM of 2.7 μm ×4.9 μm. About 4~5 J energy is deposited on the targets in each shot corresponding to a peak intensity of $2\times10^{20}$ W/cm$^2$.

The target foils are made of diamond-like carbon. Such material is thermostable up to 1100°C in vacuum and is therefore a suitable target material for laser pulses at high intensity. Under irradiation by amplified spontaneous emission (ASE) at a level of $10^8$ W/cm$^2$, the foils heat up to several hundred degrees, resulting in desorption of contaminant atoms (see supplementary information). In double foil targets, the released atoms are trapped between the foils and eventually ionized, representing dilute plasma when the main pulse arrives.

The multi-angle spectrometer is composed of a 10,000 groove/mm SiN transmission grating and a 1,000 groove/mm gold transmission grating array. The SiN grating was positioned along the laser axis, 93 cm away from the targets behind a vertical slit, covering the spectral range between 3 nm and 40 nm. The gold grating array was positioned slightly above the SiN grating to cover a spectral range between 30nm and 180nm. From the off-axis gold grating in the array, spectra under 1.3° could be registered. Behind the gratings, radiation was converted into electron signals by means of a micro-channel plate (MCP), producing visible light on a fluorescent screen which was imaged onto an ANDOR charge-coupled device (CCD) camera. By calibrating the quantum efficiency of the SiN grating and the MCP (see supplementary information), the spectra shown in figure 1 were obtained.

The intensity of the emission is calibrated by the carbon Ly-α line emission (340eV-400eV obtained for the normal double foil (NDF). For this line the conversion efficiency from laser to Ly- α line emission for a 20 nm carbon foil is already known[21] as $10^{-4}$, and its divergence is 4π.

The beam divergence of ODF emission is derived from the measured ratio $I_{1.3°}/I_{0°}$, which monotonously increases from 0.65 to 0.78 with the increase of wavelength from 50 nm to 75 nm. This collimation is not due to the cone-like profile of the synchrotron emission, as the γ here is small and the frequency is much lower than the critical frequency. Instead, it's mainly because of the optical diffraction when the source has a limited size. As a good approximation, one can use an Airy pattern to fit the measured value of $I_{1.3°}/I_{0°}$ to find out the transverse size to which the phase of the emission is locked. It was found that a circular source with diameter of 0.9 fits the measured value best. Accordingly, the divergence (full width at half maximum) of the emission at 50 nm can be extrapolated as 3.2° (56 mrad) from the Airy pattern.

Due the diffraction-induced transformation, the half-cycle feature of the field can be maintained only over a distance of $L_d \sim d^2/c\tau$, where *d* is the diameter of the source and *τ* is the width of the half-cycle pulse. In the limit of a zero-thickness FRES, theory predicts that the width of the pulse in our case is $mc/(\gamma_z eE_s) \approx 40$ as, where $E_s = n_e ed/\varepsilon_0$ is the electrostatic field created by the FRES areal charge density $n_e ed$. Plugging in this values, $L_d$~670 µm is obtained. In reality, the pulse width will be larger than this theoretical value depending on the temporal profile of the FRES. The pulse energy of ODF emission is calculated as 15 µJ, 3x10$^{-5}$ of the initial laser energy, by integrating the intensity over the spectrum (30eV-280eV) and simply assuming the beam divergence over the whole spectra is 56 mrad. It should be noticed that this value is just a rough estimate since the beam divergence at shorter wavelength (< 50 nm), which is probably slightly smaller, was not measured in the experiment.

**Simulations.** The 2D simulations were carried out with the KLAP code[22]. The simulation box has a size of 80$\lambda_L$ ×40$\lambda_L$ in the y-z plane with a resolution of 100 cells per laser wavelength $\lambda_L$. The two oblique foils are located at z=5$\lambda_L$ and z=37$\lambda_L$ on the central line at y=0, respectively. The first foil is assumed to have an exponential density profile peaked at 50 $n_c$ and a scale length of 50 nm. The second foil is chosen with a 200 $n_c$ × 50 nm step-like density profile. The gap between the two foils is filled with dilute plasma at an electron density of 0.1 $n_c$. For dilute plasma and second foil, 200 and 40 macroparticles per cell were used, respectively. The laser pulse has a peak intensity of 2.1×10$^{20}$W/cm$^2$ with a Gaussian shape in space and time. Its full-width-at-half-maximum (FWHM) duration is 50 fs. The focal spot of 4 µm FWHM-diameter is positioned at the second foil.

**Coherent emission from a kicked FRES.** Interaction with the reflected driving pulse forces transverse momenta of each FRES electron to $p_x \approx 0$ and $p_y \approx 1.1\ mc$, as seen in the simulated results (Fig. 1f) and derived analytically in the supplementary materials. Most of the switch occurs over a very short distance of $\lambda/4$ and causes radiation emission. The switch in $p_x$ is enforced by conservation of canonical momentum ($p_x - a = 0$), and the kick in $p_y$ triggers the half-cycle XUV emission that may be viewed as synchrotron motion with radius$\frac{\lambda}{4} * p_y/p_z$. Here $p_z \approx 10\ mc$ is the longitudinal momentum. The corresponding synchrotron spectrum of a single electron at $\theta = 0$ is given by[23]:

$$I_1(\omega) = \frac{d^2 I}{d\omega d\Omega} = \frac{e^2}{12\epsilon_0 \pi^3 c \gamma^4}\left(\frac{\omega\rho}{c}\right)^2 K_{2/3}^2\left(\frac{\omega\rho}{3c\gamma^3}\right) \quad (1)$$

The emission from an electron bunch at $\theta = 0$ can be described as :

$$I(\omega) = [N + N^2 F(\omega)]I_1 \quad (2)$$

Here the form factor $F(\omega) = |\int f(x)e^{-i\omega x/c}dx|^2$ is the square amplitude of the Fourier transform of the normalized electron distribution $f(x)$. The radiation is said to be coherent if the second term in equation (2) dominates. This typically occurs when the electron bunch length is comparable to the radiation wavelength, or the bunch has a very steep edge. In this case, electrons radiates in phase and the radiation is amplified by a factor of *NF(ω)* compared to incoherent emission. It is found that the FRES obtained by the simulation contains about ~10$^{10}$

(400 pC) electrons and that the form factor amounts to $10^{-3}$-$10^{-5}$ in the frequency range resolved (20 eV-80 eV). This may explain the high intensity observed. A better fit is done in figure 3 by using a tentative density profile with charge of 1000 pC and γ=15.

AUTHOR CONTRIBUTIONS:

W.J.M, J.S., J.M., and M.Z. planned the experiments. W.J.M and J.M. and M.Z. prepared the initial manuscript. H.Y.W and X.Q.Y performed the PIC simulations. All authors else contributed to the implementation of the experiments.

ACKNOWLEDGEMENTS:

The authors acknowledge the excellent support of the Astra Gemini staff, the help from Dr. Mattias Kling for providing the silicon transmission grating, and the work of the target fabrication lab of LMU.